# Retrieval Augmented Classification for Confidential Documents

**Yeseul E. Chang[1], Rahul Kailasa[2], Simon Shim[2], Byunghoon Oh[1] and Jaewoo Lee[3]**
[1]Department of Security Convergence, Chung-Ang University, Seoul, South Korea
[2]Department of Applied Data Science, San Jose State University, San Jose, CA, United States
[3]Department of Industrial Security, Chung-Ang University, Seoul, South Korea
[e-mail: yes961002@cau.ac.kr, rahul.kailasa@sjsu.edu, simon.shim@sjsu.edu, danny9807@cau.ac.kr jaewoolee@cau.ac.kr]
*Corresponding author: Jaewoo Lee

*Abstract*

Unauthorized disclosure of confidential documents demands robust, low-leakage classification. In real work environments, there is a lot of inflow and outflow of documents. To continuously update knowledge, we propose a methodology for classifying confidential documents using Retrieval Augmented Classification (RAC). To confirm this effectiveness, we compare RAC and supervised fine-tuning (FT) on the WikiLeaks US Diplomacy corpus under realistic sequence-length constraints. On balanced data, RAC matches FT. On unbalanced data, RAC is more stable while delivering comparable performance—about 96% Accuracy on both the original (unbalanced) and augmented (balanced) sets, and up to 94% F1 with proper prompting—whereas FT attains 90% F1 trained on the augmented, balanced set but drops to 88% F1 trained on the original, unbalanced set. When robust augmentation is infeasible, RAC provides a practical, security-preserving path to strong classification by keeping sensitive content out of model weights and under your control, and it remains robust as real-world conditions change in class balance, data, context length, or governance requirements. Because RAC grounds decisions in an external vector store with similarity matching, it is less sensitive to label skew, reduces parameter-level leakage, and can incorporate new data immediately via reindexing—a difficult step for FT, which typically requires retraining. The contributions of this paper are threefold: first, a RAC-based classification pipeline and evaluation recipe; second, a controlled study that isolates class-imbalance and context-length effects for FT versus RAC in confidential-document grading; and third, actionable guidance on RAC design patterns for governed deployments.

## 1. Introduction

The leakage of classified documents can cause substantial damage to nations and industries. In just the first half of 2025, they affected over 114 million victims, and this scale of damage has been growing every year [1]. As a root cause of data breaches, malicious insider attacks are infrequent but, when they occur, they have destructive power with a loss close to USD 5 million [2]. To mitigate such threats, organizations employ document classification systems that define access levels and handling procedures for sensitive information. Many countries have established security classification policies and standards (e.g., U.S. Executive Order 13526, EU Council Decision 2013/488, UK Government Security Classifications, ISO/IEC 27001/27002). However, requiring users to manually label each document's confidentiality level is labor-intensive, disrupts work continuity, and often results in inconsistent or subjective labeling [3].

Automating confidential document classification is challenging due to data scarcity and the high cost of expert labeling [4]. Traditional supervised classifiers struggle because few publicly available datasets exist, and models fine-tuned from large language models (LLMs) face generalization issues [5][6]. Fine-tuning LLMs can suffer from class imbalance (under-representation of certain confidentiality levels) and varying document lengths, which limit performance [7]. Moreover, fine-tuned models embed sensitive information in their weights, raising security concerns [8], and require costly retraining to accommodate new documents [9]. These limitations hinder the reliability and agility of deployed systems.

This study proposes a method, applying Retrieval-Augmented Classification (RAC), an efficient classification approach based on LLM [10], to classify confidential documents. This approach demonstrates its ability to address operational and management challenges arising in real world industries. This approach demonstrates superiority over existing classification methods in terms of classification accuracy, robustness. The main contributions are as follows:

- We design a confidential-document classification pipeline tailored for governed deployment and distill it.

- We empirically show that, relative to fine-tuning (FT) baselines, RAC is less sensitive to class imbalance and sequence-length constraints.

- We provide actionable guidance on RAC design patterns for governed deployments.

## 2. Related Work

### 2.1 Secret Document Classification

Recent studies have focused on the imbalance problem in classified documents [11]. To protect information assets while enabling data utilization, systematic grading that differentiates access permissions according to impact factors such as data novelty and value-creation potential is considered [12]. Han. et al. [13] proposed a prompting technique to expand the "Secret" class and demonstrated fine-tuning classification performance on data with a 2K sequence-length limitation. Bass et al. [14] introduced a method for constructing an information-security grade classification dataset and announced plans to release the resulting dataset (DISC). Nonetheless, no studies have specifically explored classification models and approaches tailored for confidential documents.

### 2.2 Retrieval-Augmented Classification

Retrieval-Augmented Classification (RAC) based on Retrieval-Augmented Generation (RAG) [9], utilizes external knowledge/examples to retrieve knowledge or similar examples not directly learned by the model and injects them into the classification process [15]. Recently, it has been shown to improve classification in environments with limited training data or class imbalance [16][17]. However, previous applications of RAC focused on general or public data, and no cases have been reported of applying this retrieval-augmented classification technique to specialized domains such as confidential documents, where security and accessibility are strictly restricted. Therefore, introducing and optimizing Retrieval-Augmented Classification techniques for specialized confidential document classification, which requires both security and accuracy, is a critical area of research.

## 3. Methodology

This study describes a RAC pipeline designed for efficient classification of classified documents, as shown in **Fig 1**. While fine-tuned models have shown effectiveness in text classification, applying them to sensitive data presents practical challenges such as class imbalance, information leakage, and retraining cost. To address these challenges, our method leverages a Retrieval-Augmented Classification (RAC) pipeline that retrieves relevant examples dynamically instead of parameter training. We use the WikiLeaks diplomatic cables dataset, which mitigates class imbalance through data augmentation. To classify classified documents, we design a multi-stage RAC system that combines advanced embedding and reranking techniques with few-shot prompting, utilizing a Large Language

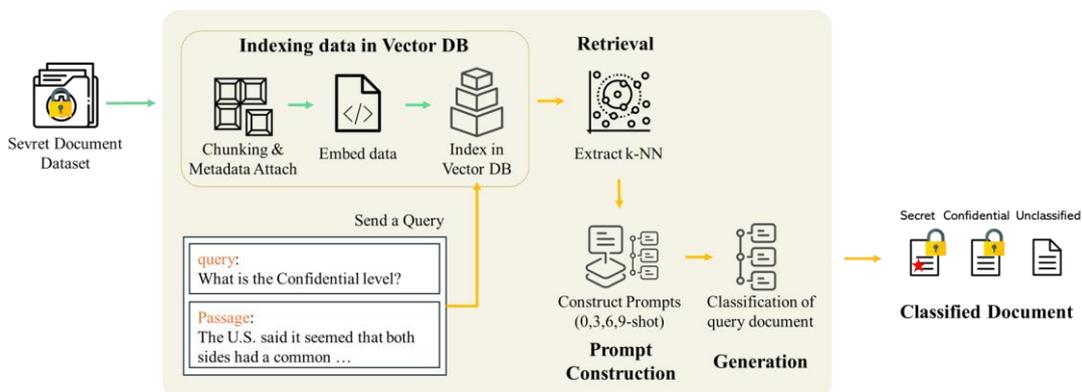

**Fig. 1.** Overview of the proposed RAC based pipeline for Confidential-document labeling

Model(LLM) for final classification. Finally, this study defines evaluation metrics such as accuracy and F1 score, and present comparisons used to verify the performance improvement of the RAC methodology compared to the baseline methodology.

### 3.1 Dataset

The WikiLeaks dataset used in this study consists of paragraph-by-paragraph transcripts of US diplomatic cables from 2003 to 2010, and is available in HTML format through a website database [18]. Each document contains metadata including title, release date, sender, recipient, content, and classification level. In this research we consolidated document label classes into standard categories: "Unclassified", "Confidential", and "Secret". This dataset was selected because it realistically represents the class imbalance and sensitive text features found in actual document archives. This study used a refined version of the WikiLeaks dataset available over Kaggle [19]. The dataset was partitioned into Training: 6,033 originals + 1,596 synthetic "secret" documents for class balancing, Test: 2,972 original cables only (no synthetic).

### 3.2 Classification with RAC

**Fig. 1** presents a RAC-based pipeline that combines dense vector retrieval with LLM inference for confidential documents classification. Training documents are embedded in batches of 64 and indexed in ChromaDB [20], which uses cosine distance with an HNSW index [21]; each record stores metadata—classification label, provenance attributes, document-length statistics, and source information—to enable quality-aware few-shot selection. Embedding representations are produced with BAAI/bge-m3 (1,024-dim) [22]; to align retrieval roles, we prefix passage: for training/indexing and query: for test/query documents, and all vectors are unit-normalized. At search time, similarity is computed as $1 -$ distance per the Chroma configuration. For each query, the system retrieves top-k = 30 similar items, reranks them with BAAI/bge-reranker-base [22], and filters by threshold. From the top results, we first select one document per class to build a 3-shot balanced prompt. If a class is missing, we issue additional retrieval to compensate and preserve balance. With these retrieved exemplars, the prompt is constructed, which prioritizes schema-based cues, official label definitions, balanced few-shot examples, and explicit decision rules for a fixed output format. Final classification is performed by OpenAI GPT-4.1, with comparative runs using meta-llama/Meta-Llama-3.1-8B-Instruct. This RAC pipeline reduces sensitivity to global label priors and avoids embedding confidential data into model parameters, enabling fast updates through re-indexing rather than retraining.

## 4. Experimental Results

To our knowledge, this is the first empirical study applying RAC to a confidential document dataset. Our results show three contributions. First, it improves Accuracy/F1 over an LLM-only classifier by leveraging retrieved evidence with balanced few-shot prompts. Second, matches or exceeds a fine-tuning(FT) baseline while avoiding retraining. Third, RAC reduces end-to-end classification time by replacing expensive training with lightweight retrieval, reranking, prompt construction, and single-pass LLM inference. Together, these results position RAC as a practical, robust, and time-efficient approach for confidential document classification.

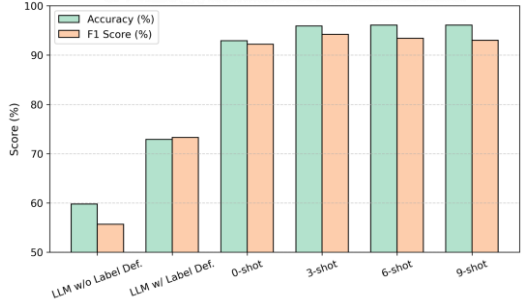

**Fig. 2.** Accuracy and F1 score of Different Strategies on the Original Imbalanced Dataset

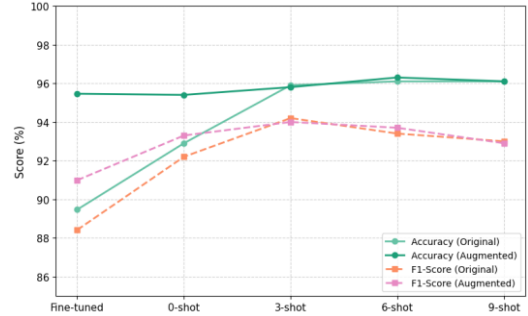

**Fig. 3.** RAC vs. FT: Performance Comparison on Original(Imbalanced) and Augmented(Balanced) Datasets

### 4.1 Experimental Setup

All experiments ran on a single NVIDIA A100-SXM4-40GB with PyTorch 2.8.0 (CUDA 12.6) and Unsloth optimizations [23]. GPT-4.1 was used for data augmentation and RAC (retrieval-augmented) classification to support large few-shot contexts, Llama-3.1-8B was used for comparable baseline, FT. This is because most of the training files are diplomatic documents containing names, organizations, and place names that are not de-identified, which resulted in many samples violating OpenAI's policies and being blocked at the upload stage. However, we mitigated this by using the same 32k token context length for LLaMA-8B to ensure a fair comparison focused on the methods rather than model size. To mitigate the minority Secret class, we generated 1,596 synthetic Secret samples via an 8-shot sliding-window prompt, enforcing semantic, lexical, and deduplication constraints. For fine-tuning, we evaluated sequence-length effects at 2,048 vs 32,768 tokens and adopted 32K for main runs to avoid truncation in long diplomatic cables. All artifacts (prompts, configs, and results) are provided in the accompanying repository. For statistical analysis of this study, we estimate 95% confidence intervals for Macro-F1 using stratified bootstrap with two thousand resamples, and compute p values with two-sided permutation tests on the same test set in a paired design. Unless stated otherwise, FT Augmented serves as the primary baseline, and the significance threshold is 0.05.

### 4.2 Evaluation Metrics

To systematically evaluate the effectiveness of the RAC-based classified confidential document classification approach, we verify classification performance. Classification performance is evaluated using accuracy and F1 score, and we compare classification performance for the classified document classification task based on two factors. We compare the performance differences between the original confidential document dataset and the augmented dataset, and we verify whether applying RAC to LLM increases classification performance. To evaluate the effectiveness of various few-shot methods in the classification model, we set the number of shots to 0, 3, 6, and 9. For comparison, we chose augmentation based Fine-Tuning as the baseline methodology for comparison.

### 4.3 Effect of RAC on Classification Performance

Applying RAC markedly improved classification over LLM-only baselines. Without label definitions, the LLM achieved 59.8% accuracy and 55.7% F1; adding label definitions raised performance to 72.9% accuracy and 73.3% F1. Introducing retrieval without examples (0-shot) further increased results to 92.9% accuracy and 92.2% F1, indicating that access to retrieved, in-distribution evidence is the main driver of gains. With balanced few-shot prompting, the 3-shot configuration yielded the best overall performance at 95.9% accuracy and 94.2% F1. Increasing shots to 6 and 9 maintained accuracy near 96.1% but reduced F1 to ~93%, suggesting diminishing returns once strong class-balanced exemplars are present and that larger prompts may introduce marginal or noisy context that slightly degrades precision–recall balance.

### 4.4 Comparison with Fine-Tuning Baseline

Fig 3 plots accuracy and F1 for fine-tuning and RAC across shot counts on the augmented and original datasets. After class balancing, the fine-tuned baseline reaches 95.46% accuracy and 90.99% F1. RAC with a zero-shot prompt attains 95.4% accuracy and 93.3% F1, while RAC with three balanced exemplars yields 95.9% accuracy and 94.2% F1. With 6-shots, accuracy peaks at 96.3% and F1 declines to 93.8% with nine shots, accuracy remains 96.1% and F1 drops to 92.9%. On the original imbalanced class, fine-tuning falls to 89.46% accuracy and 88.42% F1, whereas RAC maintains strong results, with zero-shot at 92.9% and 92.2% and 3-shot at 95.9 % and 94.2%. These results show limited dependence of RAC on augmentation. RAC provides high accuracy and matches or surpasses fine-tuning without retraining and remains robust to class imbalance. However, using too many exemplars overshooting can actually reduce minority-class F1, whereas a balanced prompt with three carefully chosen examples yields the best stability and class balance.

The RAC Original dataset with k = 0 shows statistically significant improvement over the FT Augmented dataset, increasing Macro-F1 from 0.9099 to 0.9255, with p = 9.83 × $10^{-8}$. Other RAC settings show higher Macro-F1 values, close to 0.94, but are not statistically significant (p-values > 0.12) compared to FT Augmented. The search depth tends to decrease over time, with the optimal results being achieved when k = 3 for the original data and k = 6 for the augmented data. Detailed results of the statistical tests are presented in Appendix 1.

### 5. Discussion

Our experiments show that RAC achieves comparable accuracy to fine-tuning on balanced data and better stability on imbalanced data. The first practical insight is that RAC is a safer choice, when data is severely imbalanced or confidentiality rules prevent aggregating all data into a single model. and RAC will dynamically bring in relevant examples for each query, ensuring even rare classes are recognized. Conversely, when balanced data is abundant and model retraining is feasible, fine-tuning can be performed more simply and quickly at query time. Another insight concerns model maintenance. RAC allows for immediate updates (indexing only new data), whereas fine-tuning requires a new training cycle for updates [9]. This difference can lead to significant time savings in environments where data evolves daily.

### 6. Conclusion

RAC and FT deliver comparable performance on balanced datasets. Under class imbalance, RAC remains more stable, achieving 96% accuracy and 94% F1, while FT drops from 95% to 92% F1 between balanced and unbalanced datasets. This stability results from RAC's external vector index and similarity matching, which mitigate label bias and prevent parameter-level leakage. Operationally, RAC enables immediate updates through re-indexing, improving accuracy without retraining, unlike FT which requires additional training cycles. The proposed pipeline and controlled study provide a practical framework for secure and efficient confidential-document classification. Future work will extend RAC to additional confidential datasets and explore hybrid FT+RAC models that combine FT's long-context capability with RAC's robustness and governance. We will also investigate domain adaptation to other sensitive documents such as legal or medical records.

### Acknowledgment

This research was supported by the MSIT(Ministry of Science, ICT), Korea, under the Global Research Support Program in the Digital Field program(RS-2024-00425650) supervised by the IITP(Institute for Information Communications Technology Planning Evaluation)

# Appendix

1. Result of Statistical Tests

| Model | Macro F1 | 95% CI | p (vs FT-Orig) | p (vs FT-Aug) | p (vs RAC-k0) |
|---|---|---|---|---|---|
| FT (Original) | 0.8842 | [0.8573, 0.8993] | N/A | 0.0004 | N/A |
| FT (Augmented) | 0.9099 | [0.8871, 0.9230] | 0.0004 | N/A | N/A |
| RAC-Orig (k=0) | 0.9255 | [0.9134, 0.9369] | 0.0039 | 9.83E-08 | N/A |
| RAC-Aug (k=0) | 0.9321 | [0.9183, 0.9452] | 0.3397 | 0.1588 | 7.35E-07 |
| RAC-Orig (k=3) | 0.9327 | [0.9195, 0.9455] | 0.4175 | 0.1214 | 2.07E-05 |
| RAC-Orig (k=6) | 0.9262 | [0.9102, 0.9404] | 0.0313 | 0.7465 | 1.63E-07 |
| RAC-Orig (k=9) | 0.9262 | [0.9109, 0.9403] | 0.0313 | 0.7465 | 1.63E-07 |
| RAC-Aug (k=3) | 0.9355 | [0.9220, 0.9476] | 0.039 | 0.7527 | 6.30E-08 |
| RAC-Aug (k=6) | 0.9365 | [0.9233, 0.9498] | 0.006 | 0.9336 | 2.30E-09 |
| RAC-Aug (k=9) | 0.9324 | [0.9176, 0.9459] | 0.0294 | 0.7449 | 1.10E-07 |